\documentclass[prb,twocolumn]{revtex4-1} % PR-style and notation guide: https://d22izw7byeupn1.cloudfront.net/files/styleguide-pr.pdf
\usepackage{amsmath}
\usepackage{amssymb}
\usepackage{color}

\usepackage{tikz}
\usetikzlibrary{decorations.pathreplacing,decorations.pathmorphing,shapes,calc,fadings,shadings,positioning}
\usetikzlibrary{decorations.markings}
\usepackage{braket}
\usepackage{graphicx}
\usepackage{subfigure}
\usepackage[export]{adjustbox}
\newcommand{\Si}{\hat{\mathbf{S}}_i}
\newcommand{\Sj}{\hat{\mathbf{S}}_j}

\newcommand{\Sz}{\hat{\mathbf{S}}_0}
\newcommand{\biu}{\hat{b}_{i\uparrow}}
\newcommand{\biud}{\hat{b}_{i\uparrow}^\dagger}
\newcommand{\bid}{\hat{b}_{i\downarrow}}
\newcommand{\bidd}{\hat{b}_{i\downarrow}^\dagger}
\newcommand{\bju}{\hat{b}_{j\uparrow}}
\newcommand{\bjud}{\hat{b}_{j\uparrow}^\dagger}
\newcommand{\bjd}{\hat{b}_{j\downarrow}}
\newcommand{\bjdd}{\hat{b}_{j\downarrow}^\dagger}
\newcommand{\bzu}{\hat{b}_{0\uparrow}}
\newcommand{\bzud}{\hat{b}_{0\uparrow}^\dagger}
\newcommand{\bzd}{\hat{b}_{0\downarrow}}
\newcommand{\bzdd}{\hat{b}_{0\downarrow}^\dagger}

\newcommand{\bku}{\hat{b}_{\mathbf{k}\uparrow}}
\newcommand{\bkud}{\hat{b}_{\mathbf{k}\uparrow}^\dagger}
\newcommand{\bkd}{\hat{b}_{-\mathbf{k}\downarrow}}
\newcommand{\bkdd}{\hat{b}_{-\mathbf{k}\downarrow}^\dagger}

\newcommand{\betaku}{\hat{\beta}_{\mathbf{k}\uparrow}}
\newcommand{\betakud}{\hat{\beta}_{\mathbf{k}\uparrow}^\dagger}
\newcommand{\betakd}{\hat{\beta}_{-\mathbf{k}\downarrow}}
\newcommand{\betakdd}{\hat{\beta}_{-\mathbf{k}\downarrow}^\dagger}

\newcommand{\Aij}{\hat{A}_{ij}}
\newcommand{\Aji}{\hat{A}_{ji}}
\newcommand{\Aijd}{\hat{A}_{ij}^\dagger}

\newcommand{\Bij}{\hat{B}_{ij}}
\newcommand{\Bji}{\hat{B}_{ji}}
\newcommand{\Bijd}{\hat{B}_{ij}^\dagger}

\newcommand{\matA}{{\mathcal{A}_{ij}}}
\newcommand{\matB}{{\mathcal{B}_{ij}}}
\newcommand{\Adn}{{\mathcal{A}_{\delta_1}}}
\newcommand{\Adnn}{{\mathcal{A}_{\delta_2}}}
\newcommand{\Adi}{{\mathcal{A}_{\delta_i}}}
\newcommand{\Bdn}{{\mathcal{B}_{\delta_1}}}
\newcommand{\Bdnn}{{\mathcal{B}_{\delta_2}}}
\newcommand{\Bdi}{{\mathcal{B}_{\delta_i}}}
\newcommand{\An}{\mathcal{A}_1}
\newcommand{\Ann}{\mathcal{A}_2}
\newcommand{\Bn}{\mathcal{B}_1}
\newcommand{\Bnn}{\mathcal{B}_2}
\newcommand{\Ai}{\mathcal{A}_i}
\newcommand{\Bi}{\mathcal{B}_i}
\newcommand{\A}{\mathcal{A}}
\newcommand{\B}{\mathcal{B}}
\newcommand{\wk}{\omega_{\mathbf{k}}}
\newcommand{\wq}{\omega_{\mathbf{q}}}
\newcommand{\wkq}{\omega_{\mathbf{k}-\mathbf{q}}}
\newcommand{\gka}{{\gamma_{\mathbf{k}}^A}}
\newcommand{\gkb}{{\gamma_{\mathbf{k}}^B}}
\newcommand{\gqa}{{\gamma_{\mathbf{q}}^A}}
\newcommand{\gqb}{{\gamma_{\mathbf{q}}^B}}
\newcommand{\gkqa}{\gamma_{\mathbf{k}-\mathbf{q}}^A}
\newcommand{\gkqb}{\gamma_{\mathbf{k}-\mathbf{q}}^B}

\newcommand{\veck}{\mathbf{k}}
\newcommand{\vecd}{\boldsymbol{\delta}}

\newcommand{\vecsig}{\boldsymbol{\sigma}}
\newcommand{\vecq}{\mathbf{q}}
\newcommand{\vecri}{\mathbf{r}_i}

\newcommand{\veckp}{\mathbf{k'}}
\newcommand{\vecqp}{\mathbf{q'}}

\newcommand{\Ait}{\overset{\sim}{\mathcal{A}}_{i}^2}
\newcommand{\Bit}{\overset{\sim}{\mathcal{B}}_{i}^2}

\begin{document}

% Forfatter, tittel, dato
\author{Dag-Vidar Bauer}
\author{J. O. Fj{\ae}restad}
\affiliation{Center for Quantum Spintronics, Department of Physics, Norwegian University of Science and Technology, NO-7491 Trondheim, Norway}
\title{A Schwinger boson mean field study of the $J_1$-$J_2$ Heisenberg quantum antiferromagnet on the triangular lattice}
\date{\today}

%%%%%%%%%%%%%%%%%%%%%%%%%%%%%%%%%%%%%%%%%%%%%%%%%%%%%%%%%%%%
% Abstract
%%%%%%%%%%%%%%%%%%%%%%%%%%%%%%%%%%%%%%%%%%%%%%%%%%%%%%%%%%%%
\begin{abstract}
We use Schwinger boson mean field theory (SBMFT) to study the ground state of the spin-$S$ triangular-lattice Heisenberg model with nearest ($J_1$) and next-nearest ($J_2$) neighbor antiferromagnetic interactions. Previous work on the $S=1/2$  model leads us to consider two spin liquid Ans\"{a}tze, one symmetric and one nematic, which upon spinon condensation give magnetically ordered states with 120$^{\circ}$ order and collinear stripe order, respectively. The SBMFT contains the parameter $\kappa$, the expectation value of the number of bosons per site, which in the exact theory equals $2S$. For $\kappa=1$ there is a direct, first-order transition between the ordered states as $J_2/J_1$ increases. Motivated by arguments that in SBMFT, smaller $\kappa$ may be more appropriate for describing the $S=1/2$ case qualitatively, we find that in a $\kappa$ window around 0.6, a region with the (gapped $Z_2$) symmetric spin liquid opens up between the ordered states. As a consequence, the static structure factor has the same peak locations in the spin liquid as in the 120$^{\circ}$ ordered state, and the phase transitions into the 120$^{\circ}$ and collinear stripe ordered states are continuous and first-order, respectively.  
\end{abstract}

\maketitle

\section{Introduction}
\label{intro}

\begin{figure}
\centering
\includegraphics[]{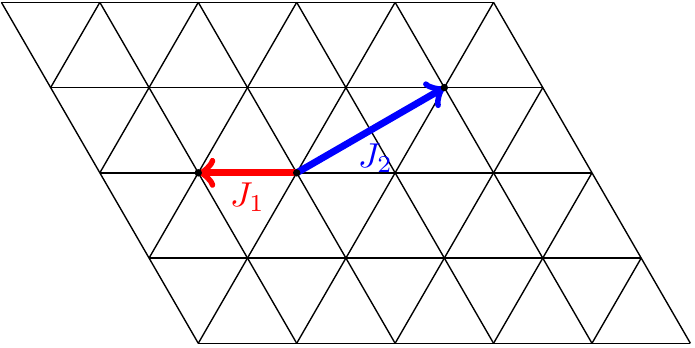}
\caption{The triangular lattice with nearest ($J_1$) and next-nearest ($J_2$) neighbour bonds indicated (for a given site there are 6 bonds of each type).}
\label{j1-j2}
\end{figure}

Quantum spin liquids have for a long time been a major research topic in frustrated quantum magnetism.\cite{bal10,mis-lhu,sav-bal} Significant progress has been made in understanding the properties of various frustrated spin systems, including the $S=1/2$ nearest-neighbour (nn) Heisenberg antiferromagnet (HAFM) on the kagome lattice, for which recent studies point towards a spin-liquid ground state\cite{norman} (but see Ref. \onlinecite{LSM-16}). A spin-liquid ground state was first proposed for the spin-$1/2$ nn-HAFM on the triangular lattice.\cite{anderson73} However, for this case it was later established that the ground state has magnetic order with a 120$^{\circ}$ angle between nn spins.\cite{trlatt-order} Here we consider the closely related $J_1$-$J_2$ HAFM on the triangular lattice, which has both nearest- ($J_1$) and next-nearest-neighbour (nnn) ($J_2$) antiferromagnetic interactions (see Fig. \ref{j1-j2}). Interest in this model was recently reinvigorated due to various numerical studies finding a spin liquid ground state in a region of intermediate values of  $J_2/J_1$.\cite{mishmash,kaneko,li,white,sheng1}

\begin{figure}
\centering
\includegraphics[]{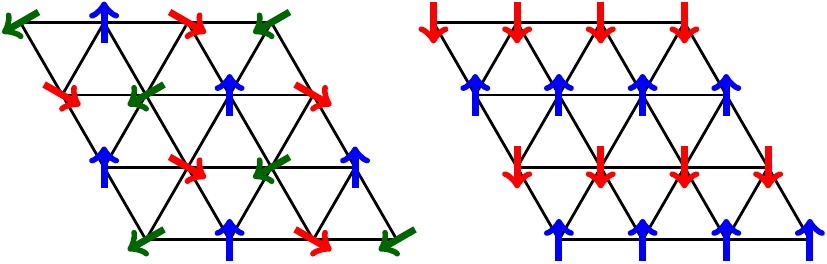}
(a) \hspace{3.5cm} (b)
\caption{Magnetically ordered phases considered in this work: (a) the $120^\circ$-ordered phase, (b) the collinear stripe phase.}
\label{ordering-patterns}
\end{figure}

We first summarize some of the main conclusions from classical (i.e. $S\to\infty$) and semiclassical analyses of the $J_1$-$J_2$ model. 
The classical model has the 120$^{\circ}$ 3-sublattice noncollinear order (Fig. \ref{ordering-patterns}a) for $J_2/J_1<1/8$, while for $1/8<J_2/J_1<1$ the ground state is characterized by a (generally) 4-sublattice order\cite{kors93} of spin vectors which sum to zero around two neighbouring elementary triangles. This leads to an infinitely degenerate ground state manifold, including both states with planar magnetic order and zero chirality,\cite{jol90} as well as states with nonplanar magnetic order and nonzero chirality\cite{kors93} (see also Ref. \onlinecite{messio11}). For $J_2/J_1>1$ a spiral state with incommensurate ordering wavevector wins out,\cite{jol90,chub92,messio11} but this region of the phase diagram will not be considered here. Spin-wave theory shows that the leading $1/S$ quantum corrections to the classical ground state lift the infinite classical degeneracy for $1/8<J_2/J_1<1$ and favor (by ''order from disorder") a state with collinear stripe order (Fig. \ref{ordering-patterns}b). This is a planar state with 2-sublattice collinear order, ferromagnetic in one direction and antiferromagnetic in the two other directions of an elementary triangle,  thus giving a 3-fold degenerate state breaking lattice rotational symmetry.\cite{jol90,kors93,chub92} Furthermore, for the $S=1/2$ case,  linear spin wave theory predicts a magnetically disordered phase in a small window around $J_2/J_1=1/8$ between the noncollinear and collinear phases.\cite{jol90,iv93,ritchey90} However, a  disordered phase was not found in nonlinear spin wave theory,\cite{chub92,deutscher93} nor in linear spin wave theory applied to finite systems.\cite{deutscher93} 

A Schwinger boson study\cite{manuel99} going beyond mean field theory found a magnetically disordered region between the two ordered phases, but did not address its nature further. More recently a number of studies\cite{mishmash,kaneko,li,white,sheng1,iqbal,sheng2,ian,lauchli,ian2,sheng3} using various numerical methods have found a spin liquid in this intermediate region, but a consensus has not yet been reached concerning its nature. All density-matrix renormalization group (DMRG) studies\cite{white,sheng1,ian,ian2,sheng3} find evidence for a gapped spin liquid, but some results that may alternatively suggest a gapless spin liquid were also found.\cite{ian,sheng3,caveats} The coupled-cluster (CCM)\cite{li} and variational Monte Carlo\cite{mishmash,kaneko,iqbal,sheng2} methods found a gapless spin liquid, with the lowest-energy state of the latter type being the U(1) Dirac spin liquid.\cite{iqbal,sheng2} The DMRG studies suggest that the spin liquid region may have nematic order, i.e. broken rotational symmetry. But these results depend on  the topological sector\cite{sheng1,ian} and could be an artifact of the explicit breaking of lattice rotation symmetry in the cylinder systems studied by DMRG. In contrast, nematic order was not found in VMC\cite{iqbal} and exact diagonalization (ED).\cite{lauchli} Ref. \onlinecite{sheng1} found evidence for a chiral spin liquid also being a possible candidate for the spin liquid in the $J_1$-$J_2$ model, but this was ruled out by later studies\cite{lauchli,sheng2,sheng3,ian2} which found that a transition to a chiral spin liquid only takes place by adding a small but finite chiral interaction term. 

Ref. \onlinecite{li} argued that the phase transitions from the spin liquid to the two ordered phases are most likely continuous. Other studies\cite{kaneko,iqbal,lauchli} agree the transition to the 120$^{\circ}$ phase is continuous but find that the transition to the collinear stripe phase is first order. Also, Refs. \onlinecite{kaneko} and \onlinecite{iqbal} calculated the static structure factor in the spin liquid region and found that the peak locations were the same as in the 120$^{\circ}$-ordered phase. There is fairly good agreement between different studies on the approximate location (in $J_2/J_1$) of the spin liquid region (see Ref. \onlinecite{iqbal} for a detailed comparison) and between DMRG and the most accurate VMC studies on the value of the ground state energy.\cite{iqbal,ian}

% Start TikZ-figur
%%%%%%%%%%%%%%%%%%%%%%%%%%%%%%%%%%%%%%%%%%%%%%%%%%%%%%%%%%%%
% Slutt TikZ-figur
%%%%%%%%%%%%%%%%%%%%%%%%%%%%%%%%%%%%%%%%%%%%%%%%%%%%%%%%%%%%

%%%%%%%%%%%%%%%%%%%%%%%%%%%%%%%%%%%%%%%%%%%%%%%%%%%%%%%%%%%%
% Tikz-figur
%%%%%%%%%%%%%%%%%%%%%%%%%%%%%%%%%%%%%%%%%%%%%%%%%%%%%%%%%%%%

%%%%%%%%%%%%%%%%%%%%%%%%%%%%%%%%%%%%%%%%%%%%%%%%%%%%%%%%%%%
% Slutt Tikz-figur
%%%%%%%%%%%%%%%%%%%%%%%%%%%%%%%%%%%%%%%%%%%%%%%%%%%%%%%%%%%

A commonly used theoretical device involves expressing the $S=1/2$ spin operator on each site in terms of either (''Abrikosov'') fermionic or (''Schwinger'') bosonic particles, whose total number on a site is fixed (the ''local constraint''). From this ''slave-particle''/''parton'' representation a mean field theory for the lattice spin model can be constructed, in which the spin-spin interactions are approximated by effective quadratic terms, and the local number operator constraints are replaced by a (weaker)  
expectation value constraint. The effective quadratic Hamiltonian is then diagonalized and its coefficients are determined self-consistently. Such mean-field  theories may give qualitative insights and/or may be a starting point for more refined methods (including the VMC studies already noted, in which the variational trial states are fermionic mean field states numerically projected to satisfy the local constraint). The bosonic formulation is in fact valid for any (integer or half-integer) $S$. In the associated (Schwinger boson) mean-field theory (SBMFT)\cite{arovas-auerbach} it is possible and useful to consider $\kappa=2S$ as a continuous parameter. A candidate mean field state will represent a gapped spin liquid for $\kappa<\kappa_c$ and a magnetically ordered state for $\kappa>\kappa_c$ (the transition occurring by Bose condensation of the bosonic ''spinon'' excitations), where $\kappa_c$ depends on the model parameters. SBMFT was applied to the $S=1/2$ $J_1$-$J_2$ model in Ref. \onlinecite{gazza93}: taking $\kappa=1$ they found a direct, first-order transition between the 120$^{\circ}$ and collinear stripe ordered states at $J_2/J_1\approx 0.16$ (see also Ref. \onlinecite{merino}). Later, by analyzing the spin stiffness, Ref. \onlinecite{manuel99} found that one-loop corrections to the mean field theory led to a small $J_2/J_1$-region with a magnetically disordered phase between the two ordered phases. 

One of the aims of our work is to investigate the nature of this disordered phase in the bosonic formulation, a task that was not undertaken in Ref. \onlinecite{manuel99}. We will however stay purely within the mean field theory. This requires some explanation since, as already noted, SBMFT for $\kappa=1$ found no disordered phase. But although $\kappa=1$ is the correct choice in an exact treatment of the $S=1/2$ model, a lower value of $\kappa$ may be more appropriate in mean field theory.\cite{messio2010,psgmessio2013} We will therefore consider the phase diagram as a function of $\kappa$ which is taken to be a free continuous parameter. Not fixing $\kappa$ also gives more insight into which states may be energetically close in parameter space and is generally more consistent with the fact that the information provided by mean field theory is at best qualitative. 

The recent numerical studies reviewed earlier suggest that the relevant spin liquid candidate states are nonchiral and may or may not be nematic. Thus it is necessary to consider symmetry properties of parton mean field states. Wen introduced the concept of the projective symmetry group (PSG) and used it to derive and classify spin liquid mean field Ans\"{a}tze in the fermionic formulation.\cite{psgwen2002prb65,wenbook} Wang and Vishwanath\cite{psgWV2006} adapted this approach to the bosonic formulation by using a PSG analysis to derive SBMFT spin liquid Ans\"{a}tze. These works considered symmetric spin liquid Ans\"{a}tze representing physical states invariant under space group transformations, spin rotations and time reversal (thus not including chiral states). For  the triangular lattice, Ref. \onlinecite{psgWV2006} found eight such Ans\"{a}tze in the bosonic formulation. These findings have been reproduced by later analyses which have also mapped the bosonic Ans{\"{a}tze to corresponding fermionic ones.\cite{qi1,qi2,lu1,lu2} However, based on the mean-field parameters they allow, only two of the eight Ans{\"{a}}tze have been considered as promising candidates for the $J_1$-$J_2$ model:\cite{psgWV2006,qi1,lu1} the 0-flux state (previously identified by  Sachdev\cite{sachdev92} in a large-${\cal N}$ bosonic formulation) and the $\pi$-flux state.\cite{terminology} A natural question is whether these two states could upon spinon condensation give rise to precisely the two types of magnetically ordered states found in the $J_1$-$J_2$ model. Indeed, this is the connection between the 0-flux state and the 120$^{\circ}$  order.\cite{sachdev92,psgWV2006} On the other hand, although the magnetic order associated with the $\pi$-flux state was found to have the same ordering wavevectors as the collinear stripe order (3 possible ordering vectors, located at the Brillouin zone edge centers), the actual magnetic order was found to be different.\cite{psgWV2006,pifluxcriticism}  

We are not aware of any PSG analysis for the SBMFT formulation that has found a symmetric spin liquid Ansatz whose associated magnetic ordering is that of the collinear stripe phase (with 3 possible ordering vectors). Thus we are led to look for an Ansatz which is not fully symmetric. As the collinear stripe state breaks the lattice rotational symmetry, it is natural to consider an Ansatz that does the same (i.e. a nematic spin liquid\cite{lu1}) and upon spinon condensation gives rise to collinear stripe order with a unique ordering vector. In this work we study the competition between the 0-flux state, the nematic spin liquid (NSL) state, and the magnetically ordered states these can give rise to. 

This paper is organized as follows: Sec. \ref{theory} discusses the SBMFT for the 0-flux and NSL states. Most of the numerical results, including the ground state phase diagram as a function of $J_2/J_1$ and $\kappa$, are presented in Sec. \ref{results}. Sec. \ref{conclusions} gives a discussion and conclusions.  Derivations of the static structure factor and a small-$\kappa$ expansion are included in two appendices.   

%%%%%%%%%%%%%%%%%%%%%%%%%%%%%%%%%%%%%%%%%%%%%%%%%%%%%%%%%%%%
% Teori
%%%%%%%%%%%%%%%%%%%%%%%%%%%%%%%%%%%%%%%%%%%%%%%%%%%%%%%%%%%%

\section{Theory}
\label{theory}

% Avsnitt: SBMFT
%%%%%%%%%%%%%%%%%%%%%%%%%%%%%%%%%%%%%%%%%%%%%%%%%%%%%%%%%%%%
\subsection{Schwinger boson mean field theory}

We will investigate the $J_1$-$J_2$ HAFM on the triangular lattice (Fig. \ref{j1-j2}). The Hamiltonian is
\begin{equation}
  \label{heisenberg-modellen} 
  H=J_1\sum_{\langle i,j\rangle} \Si \cdot \Sj + J_2\sum_{\langle\langle i,j \rangle\rangle} \Si \cdot \Sj
\end{equation}
where the sums run over pairs of nn and nnn sites, respectively, each pair being counted once. Periodic boundary conditions will be imposed on the spins.

In the Schwinger boson representation the spin operators $\Si$ are written as 
\begin{equation} 
  \Si = \frac{1}{2}\sum_{\alpha\beta}\hat{b}_{i\alpha}^\dagger \sigma_{\alpha\beta} \hat{b}_{i\beta}, 
\end{equation}
where $\vecsig$ is the vector of Pauli matrices and $\hat{b}_{i\sigma}^\dagger$ and $\hat{b}_{i\sigma}$ are creation and annihilation operators for a boson with spin $\sigma=\{\uparrow,\downarrow\}$ on lattice site $i$; these operators satisfy the
standard commutation relations $ [\hat{b}_{i\alpha},\hat{b}^{\dagger}_{j\beta}]=\delta_{ij}\delta_{\alpha\beta}$. To 
enforce that $\Si^2 = S(S+1)$, the operator identity
\begin{equation}
  \label{constraint}
  \hat{n}_i = \sum_{\sigma} \hat{b}_{i\sigma}^\dagger \hat{b}_{i\sigma} = 2S
\end{equation}
should hold at each site $i$; this is the local constraint.

As spin liquid states don't break spin rotation symmetry, in SBMFT 
one seeks to express the Hamiltonian in terms of quadratic operators that don't break this symmetry, letting the expectation value of these operators serve as mean-field parameters. The only quadratic operators that qualify are
\begin{subequations}
  \begin{align} 
    &\Aij = \frac{1}{2}(\biu\bjd-\bid\bju),\\ 
    &\Bij = \frac{1}{2}(\biu\bjud+\bid\bjdd),
  \end{align}
\end{subequations}
and their adjoints. These ''bond operators'' satisfy $\Aij=-\Aji$ and $\Bijd=\Bji$.  
The Heisenberg interaction can then be written  
\begin{subequations}
  \begin{align} 
    \Si\cdot\Sj 
    &= :\Bijd\Bij:-\Aijd\Aij\\ 
    &= \Bijd\Bij-\Aijd\Aij-\frac{1}{4}\hat{n}_i, 
  \end{align}
\end{subequations}
where $:\_:$ means normal-ordering.

In the mean-field approximation, we write $\Aij = \langle \Aij \rangle + (\Aij-\langle \Aij \rangle) \equiv \langle \Aij \rangle + \delta\Aij$, and similarly for the field $\Bij$. Ignoring deviations from the mean of order $(\delta A)^2$, we obtain the $A$-term
\begin{equation} 
  \Aijd\Aij \simeq \braket{\Aij}^*\Aij+\Aijd\braket{\Aij}-|\braket{\Aij}|^2,
\end{equation}
and similarly for the $B$-term. The mean-field parameters $\braket{\Aij}$,$\braket{\Bij}$ will from now on be denoted by $\matA,\matB$. The set of mean-field parameters $\{\matA,\matB\}$ is referred to as an Ansatz. 

In SBMFT, the local constraint \eqref{constraint} is relaxed to hold only at the level of expectation values, i.e. 
\begin{equation} 
  \braket{n_i} = \kappa.
\label{mf-constraint}
\end{equation}
(Although naively $\kappa=2S$, other choices of $\kappa$ can be justified, as discussed later.)  Thus we should
minimize the mean-field Hamiltonian $\hat{H}_{\text{MF}}$ with respect to $\{\matA,\matB\}$, subject to the $N$ local constraints (\ref{mf-constraint}) ($N$ is the total number of sites). This is done by adding to $\hat{H}_\text{MF}$ a term $\sum_i \lambda_i(\hat{n}_i-\kappa)$ where $\{\lambda_i\}$ is a set of Lagrange multipliers. This gives 
\begin{align}
\label{H_MF}
\hat{H}_\text{MF} &= \sum_i \lambda_i(\hat{n}_i-\kappa) + \left(J_1\sum_{\langle i,j \rangle} + J_2 \sum_{\langle\langle i,j \rangle\rangle}\right) \notag \\
&  \hspace{-1cm}\left\{(\Bijd\matB - \Aijd\matA + \mbox{h.c.})+|\matA|^2-|\matB|^2-\frac{1}{4}\hat{n}_i\right\}.
\end{align}
%
%\begin{align}
%  \label{H_MF} 
%  \hat{H}_\text{MF} &= J_1\sum_{\langle i,j \rangle} \left(\matB\Bij + \Bijd\matB-\matA\Aij - \Aijd\matA\right) \notag \\ 
%  &+ J_2\sum_{\langle\langle i,j \rangle\rangle} \left(\matB\Bij + \Bijd\matB-\matA\Aij - \Aijd\matA\right) \notag \\ 
%  &+ J_1\sum_{\langle i,j \rangle}\left(\matA^2-\matB^2-\frac{1}{4}\hat{n}_i\right) \notag\\ 
%  &+J_2\sum_{\langle\langle i,j \rangle\rangle}\left(\matA^2-\matB^2-\frac{1}{4}\hat{n}_i\right)\notag\\ 
%  &+\sum_i \lambda(\hat{n}_i-\kappa). 
%\end{align}
%

\subsection{States and Ans\"{a}tze}
\label{states-and-ansatze}

As discussed in Sec. \ref{intro}, motivated by previous work we are led to consider the competition between two spin liquid states: the 0-flux state and a nematic spin liquid (NSL) state, whose magnetic ordering patterns are the 120-degree order and a collinear stripe order, respectively (see Fig. \ref{ordering-patterns}), both of the coplanar type. As these states are nonchiral, they can be described by real Ans\"{a}tze.\cite{psgmessio2013} Also, these states' Ans\"{a}tze have the same translation symmetry as the lattice, so the Ansatz unit cell consists of a single site. The mean-field parameters $\matA$, $\matB$ can therefore only depend on $\bf{r}_i-\bf{r}_j$. Only the nn parameters (denoted by $\Adn,\Bdn$) and nnn parameters (denoted by $\Adnn,\Bdnn$) will enter into the determination of the mean-field solution.  These parameters are listed for the two states in Table \ref{ansatztable} (see also Fig. \ref{reference-directions}).

\begin{figure}\centering
\includegraphics[]{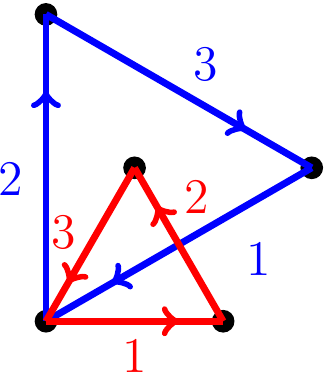}
\caption{Nearest-neighbor (red) and next-nearest-neighbor (blue) bonds used for specifying Ans\"{a}tze. The arrows indicate our choice of positive reference directions for $\mathcal{A}$-parameters (an arrow from site $i$ to site $j$ means $\mathcal{A}_{ij}>0$). The oriented nn bonds 1 and 2 also define the basis vectors $\mathbf{a}_1$ and $\mathbf{a}_2$ for the triangular lattice.}
\label{reference-directions}
\end{figure}

% Tabell over symmetrier
\begin{table}[h!]\centering
  \begin{tabular}{ccc}
    \hline
    \multicolumn{2}{r}{States} \\
    \cline{2-3}
    Parameters & 0-flux & NSL \\
    \hline
    $\Adn$  & $(\A,\A,\A)$    & $(0,\A,\A)$        \\
    $\Bdn$  & $-(\B,\B,\B)$  & $-(\B,0,0)$       \\
    $\Adnn$ & $(0,0,0)$     & $(\bar{\A},0,\bar{\A})$      \\
    $\Bdnn$ & $(\bar{\B},\bar{\B},\bar{\B})$ & $(0,\bar{\B},0)$  \\
    \hline
  \end{tabular}
\caption{Nearest- ($\delta_1$) and next-nearest ($\delta_2$) neighbor mean field parameters for the 0-flux and nematic spin liquid (NSL) states studied in this work. The components of the triples are bond parameters for (nn or nnn) bonds 1-3 in Fig. \ref{reference-directions}.}
\label{ansatztable}
\end{table}

The $0$-flux state is characterized by having equal magnitude for all 
nn $\Adn$, $\Bdn$ and nnn $\Bdnn$, while the nnn $\Adnn$ vanish. \cite{psgWV2006} The name ''0-flux'' derives from
the gauge-invariant flux $\Phi \equiv \mathrm{arg}(\mathcal{A}_{ij}\mathcal{A}_{jk}\mathcal{A}_{kl}\mathcal{A}_{li})=0$ around a rhombus.  

The NSL state is 3-fold degenerate, breaking lattice rotational symmetry by having ferromagnetic spin correlations along one of the three directions of a triangle and antiferromagnetic correlations along the two other  directions. The parameters shown in Table \ref{ansatztable} correspond to ferromagnetic correlations in the horizontal direction, cf. Fig. \ref{ordering-patterns}b. 

%%%%%%%%%%%%%%%%%%%%%%%%%%%%%
% Dispersjons-plott (SETT INN RIKTIGE PLOT)
%%%%%%%%%%%%%%%%%%%%%%%%%%%%
\begin{figure*}[htb]
\centering
{\includegraphics[width=0.3\textwidth,valign=c]{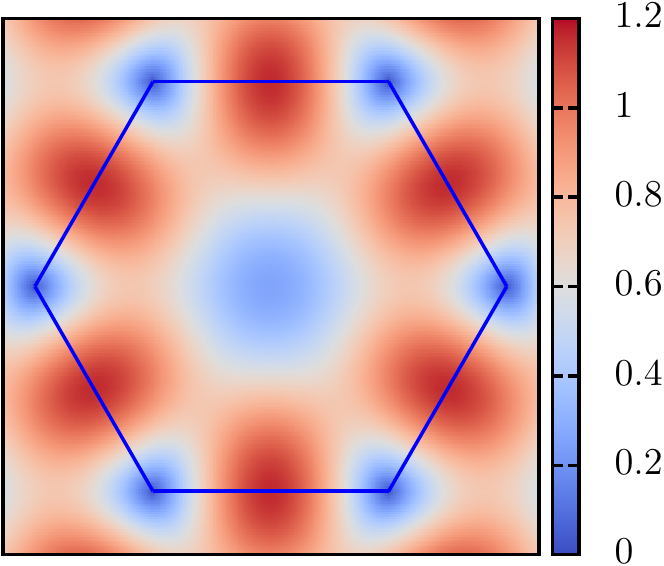}}
\hspace{0.12cm}
{\includegraphics[width=0.3\textwidth,valign=c]{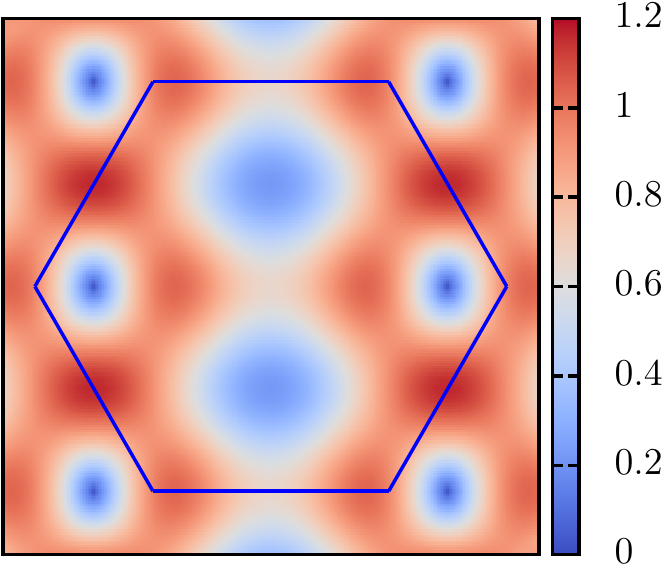}}
\hspace{0.12cm}
{\includegraphics[width=0.25\textwidth,valign=c]{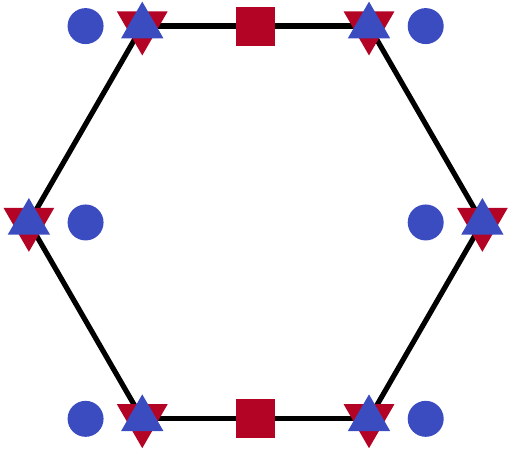}}
\vspace{0.12cm}
  \caption{Left and middle: Spinon dispersion $\omega_{\mathbf{k}}$ for the two states evaluated for $\kappa=1$ and $L=48$  (left: 0-flux state for $J_2/J_1=0$, middle: NSL state for $J_2/J_1=0.5$). In both plots, $\kappa>\kappa_c(J_2/J_1)$, so magnetic order is present. The plots of $\omega_{\mathbf{k}}$ look qualitatively the same also for $\kappa<\kappa_c$, but the variations are smoother. Right: Locations of the spinon dispersion minima in the 0-flux state (blue up-pointing triangles) and in the NSL state (blue circles), locations of ordering vectors in the 0-flux state (red down-pointing triangles) and in the NSL state (red squares). The hexagon is the 1st Brillouin zone of the triangular lattice.}
  \label{fig:spinon-dispersion}
\end{figure*}
%%%%%%%%%%%%%%%%%%%%%%%%%%%%%

\subsection{Solving the SBMFT}
\label{solving}

As the Ans\"{a}tze $\{\matA,\matB\}$ to be considered here are translationally invariant, we expect the Lagrange multipliers to be site-independent, so we set $\lambda_i\equiv \lambda$. The $N$ local constraints (\ref{mf-constraint}) thus reduce to a single global constraint $\sum_i \langle n_i\rangle =\kappa N$, implemented by a single Lagrange multiplier $\lambda$. By introducing a Fourier transformation $\hat{b}_{i\sigma}=\frac{1}{\sqrt{N}}\sum_{\veck}\mathrm{e}^{\mathrm{i}\veck\cdot\vecri}\hat{b}_{\veck\sigma}$, the Hamiltonian is block-diagonalized as (we redefine 
$\lambda \mapsto \lambda-\frac{1}{8}\sum_{\delta_1} J_1 - \frac{1}{8}\sum_{\delta_2} J_2$)
\begin{align} 
\lefteqn{\hat{H}_\text{MF} = \sum_{\mathbf{k}}(\gkb+\lambda)(\bkud\bku+\bkd\bkdd)} \notag \\ &+ \sum_{\mathbf{k}}\mathrm{i}\gka(\bku\bkd-\bkud\bkdd)-N\lambda(\kappa+1)\notag\\ &+\frac{N}{2}\sum_{i=1}^2\sum_{\delta_i}J_i\left(\Adi^2-\Bdi^2\right),
\end{align}
where here and in the following we have omitted a constant $C = -(N\kappa/8)\sum_{i=1}^2\sum_{\delta_i}J_i$ on the right-hand side, and we have introduced
\begin{subequations}
  \begin{align} 
    &\gka  = \frac{1}{2}\sum_{i=1}^2\sum_{\delta_i}J_i\Adi\sin(\veck\cdot\vecd_i),\\ 
    &\gkb = \frac{1}{2}\sum_{i=1}^2\sum_{\delta_i}J_i\Bdi\cos(\veck\cdot\vecd_i).
  \end{align}
\end{subequations}
The diagonalization is completed with a Bogoliubov transformation, 
\begin{subequations} \label{bogolyubov}
  \begin{align}
    &\bku=\cosh\theta_\veck \betaku-\sinh\theta_\veck \hat{\beta}_{-\veck\downarrow}^\dagger,\\ 
    &\bkdd= \mathrm{i}\sinh\theta_\veck \betaku-\mathrm{i}\cosh\theta_\veck \hat{\beta}_{-\veck\downarrow}^\dagger,
  \end{align}
\end{subequations}
with
\begin{equation}
	\tanh 2\theta_\veck = -\frac{\gka}{\gkb+\lambda}.
    \label{diagonal-condition}
\end{equation}
The result is
\begin{equation} 
\hat{H}_\text{MF} = E_0+\sum_{\mathbf{k}}\wk(\betakud\betaku+\betakdd\betakd),
\end{equation}
where the dispersion of the bosonic excitations (the ''spinons'') is
\begin{equation}
  \wk = \sqrt{(\gkb+\lambda)^2-(\gka)^2}, 
\end{equation}
and the ground-state energy $E_0$ is
\begin{align}
    E_0 &= \frac{N}{2}\sum_{i=1}^2\sum_{\delta_i} J_i\left(\Adi^2-\Bdi^2\right) \notag \\&-N(\kappa+1)\lambda+\sum_\veck \wk.
\end{align}
The mean-field parameters are determined from
\begin{equation} 
  \frac{\partial E_0}{\partial \Adi} = 0, \quad \frac{\partial E_0}{\partial \Bdi} = 0, \quad
\frac{\partial E_0}{\partial \lambda} = 0,
\end{equation}
which leads to the mean-field equations
\begin{subequations}
\label{mean-field-equations}
  \begin{align}
    &\Adi = \frac{1}{2N}\sum_\veck \frac{\gka}{\wk}\sin(\veck\cdot\vecd_i),\\ 
    & \Bdi =\frac{1}{2N}\sum_\veck \frac{\gkb+\lambda}{\wk}\sin(\veck\cdot\vecd_i),\\
    & 1+\kappa = \frac{1}{N}\sum_\veck \frac{\gkb+\lambda}{\wk}.
  \end{align}
\end{subequations}
Using the mean-field equations, %(or Eq. (\ref{H_MF}))
 the ground-state energy can be rewritten as
\begin{subequations}
  \begin{align}
    E_0 &
     = \frac{1}{2}\sum_\veck \wk -\frac{N}{2}\lambda(\kappa+1)\notag\\
    & =\frac{N}{2}\sum_{i=1}^2\sum_{\delta_i}J_i\left(\Bdi^2-\Adi^2\right).
  \end{align}
\end{subequations} 

It can be verified that the local constraints (\ref{mf-constraint}) are satisfied, as expected. Given Eq. (\ref{constraint}), setting $\kappa=2S$ seems natural, and indeed this has been a standard choice in the literature. However, it is not the only or necessarily the best choice.\cite{messio2010,psgmessio2013} It can be shown (see Appendix \ref{Sqder}) that our SBMFT gives $\braket{\Si^2}=\frac{3}{8}\kappa(\kappa+2)$, so the correct result $S(S+1)$ is overshot by a factor $3/2$ for $\kappa=2S$.\cite{auerbach-book} Choosing to solve the mean-field theory subject to the alternative constraint that $\braket{\Si^2}$ takes the correct value would give a smaller value of $\kappa$ (in particular, $S=1/2$ would give $\kappa=\sqrt{3}-1\approx 0.73$). In view of this nonuniqueness, and in order to get more insights from the mean field theory (whose conclusions are in any case at best qualitative), we will treat $\kappa$ as a continuous parameter in the theory. It can then be used to extrapolate between the extreme quantum limit ($\kappa=0$) and the classical limit ($\kappa=\infty$), and to determine the critical parameter value $\kappa_c$ below which the quantum fluctuations destroy magnetic order ($\kappa_c$ will depend on the Ansatz and $J_2/J_1$).

%%%%%%%%%%%%%%%%%%%%%%%%%%%%%%%%%%%%%
% Magnetisk orden
%%%%%%%%%%%%%%%%%%%%%%%%%%%%%%%%%%%%
\subsection{Spin correlations and magnetic order}

In order to investigate spin correlations and possible magnetic order we consider the correlation function
\begin{equation}
\braket{\Si \cdot \Sj}  = \frac{1}{N}\sum_{\vecq}S(\mathbf{q})\mathrm{e}^{\mathrm{i}\mathbf{q}\cdot(\mathbf{r}_i-\mathbf{r}_j)}
\end{equation}
where its Fourier transform $S(\mathbf{q})$, the static structure factor, is given by (see Appendix \ref{Sqder} for a derivation)
\begin{equation}
\label{S(q)} S(\mathbf{q}) = \frac{3}{8N}\sum_{\veck}\left[\frac{(\gkb+\lambda)(\gkqb+\lambda)-\gka\gkqa}{\wk\wkq}-1\right].
 \end{equation}
Maxima in $S(\mathbf{q})$ occur for $\mathbf{q}\in \{\mathbf{q}_0\}$ due to terms in (\ref{S(q)}) for which $\mathbf{k}$ and $\mathbf{k}-\mathbf{q}_0$ are inequivalent spinon dispersion minima. Plots of the spinon dispersion for the 0-flux and NSL state are given in Fig. \ref{fig:spinon-dispersion}, which also shows the locations of the spinon dispersion minima (for each state there are two such wavevectors $\pm \mathbf{k}_0$ in the BZ) and the wavevectors $\{\mathbf{q}_0\}$ of the dominant magnetic correlations. In the 0-flux state, both $\pm \mathbf{k}_0$ and $\{\mathbf{q}_0\}$ consist of the two inequivalent wavevectors $\pm\mathbf{Q}$ at the BZ corners. In the NSL state, for our choice of ferromagnetic correlations in the horizontal direction, $\pm \mathbf{k}_0=(\pm \pi,0)$, and $\{\mathbf{q}_0\}$ consists of a single vector $\mathbf{Q}$, at the middle of the horizontal BZ edge.

To investigate magnetic order, we consider the dominant contribution to $ \braket{\Sz\cdot\Si}$,
\begin{equation}
\frac{1}{N}\sum_{\vecq \in \{\mathbf{q}_0\}}S(\mathbf{q})\mathrm{e}^{\mathrm{i}\mathbf{q}\mathbf{r}_i}
=\frac{{\cal N}_{\mathbf{q}_0}S(\mathbf{Q})}{N}\cos(\mathbf{Q}\cdot\mathbf{r}_i)
\end{equation}
where ${\cal N}_{\mathbf{q}_0}$ is the number of vectors in $\{\mathbf{q}_0\}$. This motivates the definition of a sublattice magnetization parameter
\begin{equation}
m^2(N) \equiv \frac{{\cal N}_{\mathbf{q}_0}S(\mathbf{Q})}{N} \simeq \frac{3}{2}\left(\frac{\gamma_{\mathbf{k}_0}^A}{N\omega_{\mathbf{k}_0}}\right)^2,
\label{magnetization-parameter}
\end{equation}
where in the last expression $S(\mathbf{Q})$ was approximated by the biggest term(s) in (\ref{S(q)}) (in the 0-flux state, this comes from the spinon minimum at $-\mathbf{Q}$, while in the NSL state it includes both spinon minima).\cite{compare-mezio} Thus in the magnetically ordered phase, characterized by $m^2(N)$ approaching a nonzero value in the thermodynamic limit, $S(\mathbf{Q})$ diverges linearly with $N$ and the spinon dispersion minimum (spinon gap) $\omega_{\mathbf{k}_0}$ scales to 0 like $1/N$.

%%%%%%%%%%%%%%%%%%%%%%%%%%%%%%%%%%%%%%%%%%%%%%%%%%%%%%%%%%%%
% Resultater
%%%%%%%%%%%%%%%%%%%%%%%%%%%%%%%%%%%%%%%%%%%%%%%%%%%%%%%%%%%%
\section{Results}
\label{results}

We have solved the self-consistent equations \eqref{mean-field-equations} numerically for finite lattices with $N=L^2$ sites, with periodic boundary conditions after $L$ sites along the $\mathbf{a}_1$ and $\mathbf{a}_2$ directions (defined in Fig. \ref{reference-directions}). We used $L$-values in the range 12-60, further restricted by requiring that the spinon dispersion minima should lie on the numerical grid of $\mathbf{k}$-vectors (giving $L$ divisible by 6 and 4 in the 0-flux and NSL phase, respectively). Such finite-$N$ calculations can be used to
determine the lowest-energy phase in most of the $(J_2/J_1,\kappa)$ parameter space. As an example, Fig. \ref{fig:energy} shows the ground state energy for $L=48$ as a function of $J_2/J_1$ for various values of $\kappa$. For each $\kappa$, the ground state is 0-flux at small $J_2/J_1$ and NSL at larger $J_2/J_1$, with the first-order transition point $(J_2/J_1)_c$ increasing with decreasing $\kappa$. 
%
%%%%%%%%%%%%%%%%%%%%%%%%%%%%%%%%%%%%%%%%%%%%%%%%%%%%%%%%%%%%
\begin{figure}
  \begin{center}
  \includegraphics[width=0.45\textwidth]{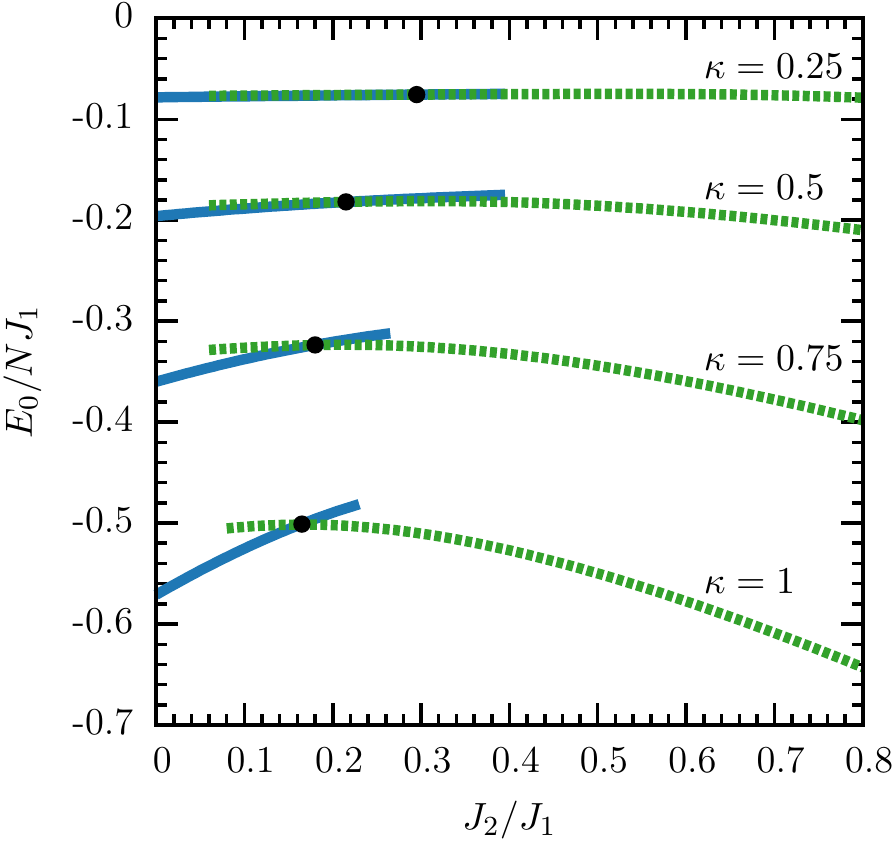}
    \caption{Ground state energies of the $0$-flux state (full lines) and the NSL state (dotted lines) for $L=48$ and various values of $\kappa$.}
    \label{fig:energy}
  \end{center}
\end{figure}
%%%%%%%%%%%%%%%%%%%%%%%%%%%%%%%%%%%%%%%%%%%%%%%%%%%%%%%%%%%%
%

In order to determine more precisely the boundary between the two phases, and the boundary between magnetic order and disorder within a given phase, we have also considered extrapolations of finite-$N$ results to the thermodynamic limit. 
By fitting $m^2(N)$ in (\ref{magnetization-parameter})\cite{fits} to the scaling form\cite{sandvik-scaling-1997} $m^2(N) = m_0^2 + a/L+b/L^2+c/L^3$, we determined the critical value $\kappa_c(J_2/J_1)$ for magnetic order from the estimated onset of a positive value of the fitting parameter $m_0^2$. We have done such fits for lattices up to $N=3600$. We have also used an alternative method in which $\kappa_c$ is found as the intersection of plots of $\xi_a/L$ for different (large) $L$, where $\xi_a$ is the correlation length measure 
$\xi_a = |\mathbf{q}_1|^{-1}\sqrt{S(\mathbf{Q})/S(\mathbf{Q}+\mathbf{q}_1)-1}$, with $\mathbf{q}_1$ the smallest nonzero wavevector along some chosen direction. \cite{sandvik-correlation-2010} To determine the ground state energy in the magnetically ordered phases in the thermodynamic limit, we used the scaling form\cite{sandvik-scaling-1997} $E_0(N) = E_0 + A/L^3 + B/L^4 + C/L^5$. In the magnetically disordered phases below $\kappa_c$, where the spinon gap $\omega_{\mathbf{k}_0}$ is finite, the energy was found to be so well converged for our largest $L$-values that further extrapolations were not needed.
\begin{figure}[!t]
\begin{center}
    \includegraphics[width=0.49\textwidth]{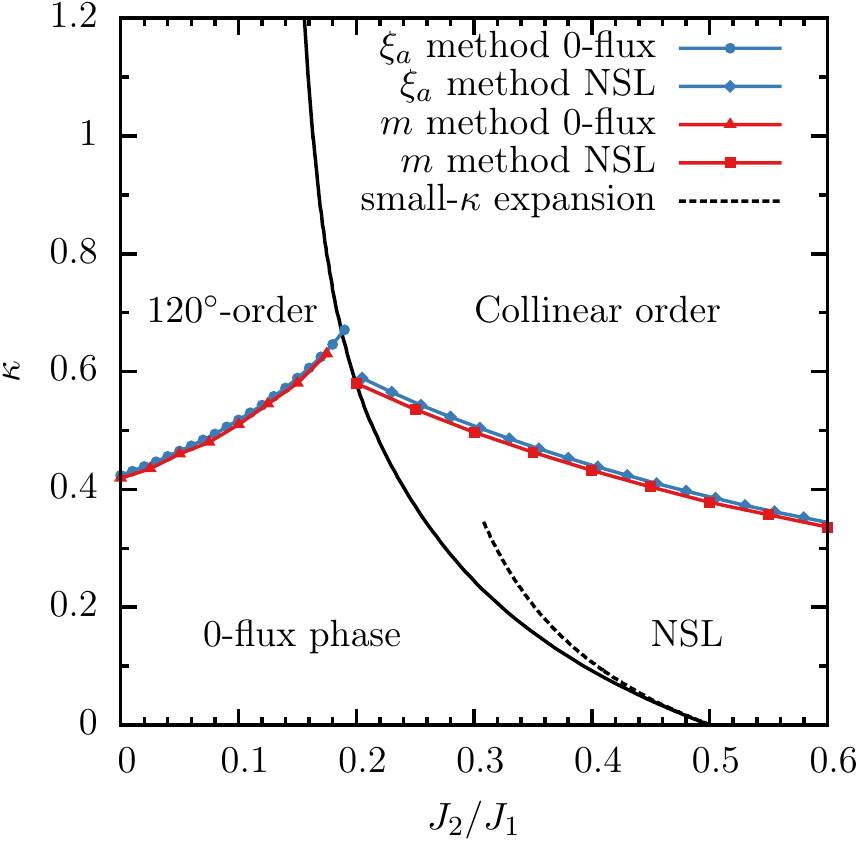}
    \caption{Zero temperature phase diagram for the $J_1$-$J_2$-model.}
    \label{fig:phasediagram}
\end{center}
\end{figure}

The ground state phase diagram resulting from the numerical calculations is shown in Fig. \ref{fig:phasediagram}. We now discuss some aspects of this phase diagram. 

The phase transition between the 0-flux and NSL phase (full line) moves to higher $J_2/J_1$ as $\kappa$ is reduced, ending at $J_2/J_1=1/2$ in the limit $\kappa\to 0$. We have also calculated the ground state energies
analytically using a small-$\kappa$ expansion\cite{smallkappa} with terms up to and including $O(\kappa^3)$ (see Appendix \ref{small-kappa}); as expected, the resulting transition (dashed line) is found to agree with the numerical curve for small enough $\kappa$.

%%%%%%%%%%%%%%%%%%%%%%%%%%%%%
% Struktur-faktor-plott
¤%%%%%%%%%%%%%%%%%%%%%%%%%%%%
\begin{figure*}[!htbp]
  \centering
  \includegraphics[width=0.3\textwidth]{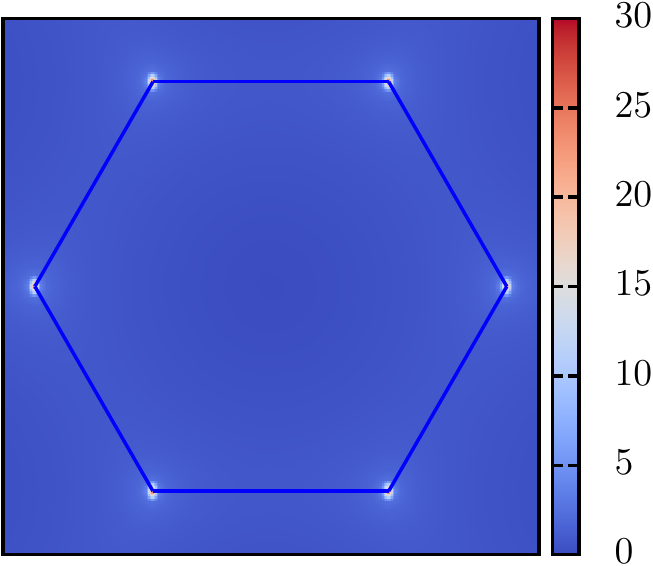} 	
  \hspace{0.1cm}
  \includegraphics[width=0.3\textwidth]{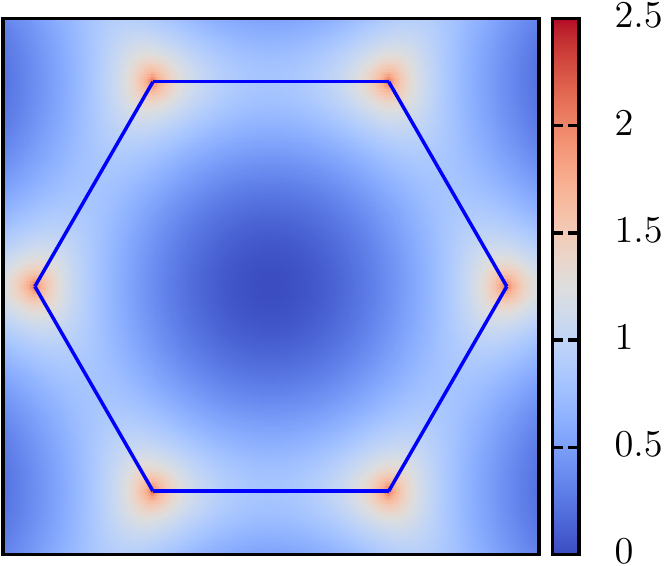}
  \hspace{0.1cm}
  \includegraphics[width=0.3\textwidth]{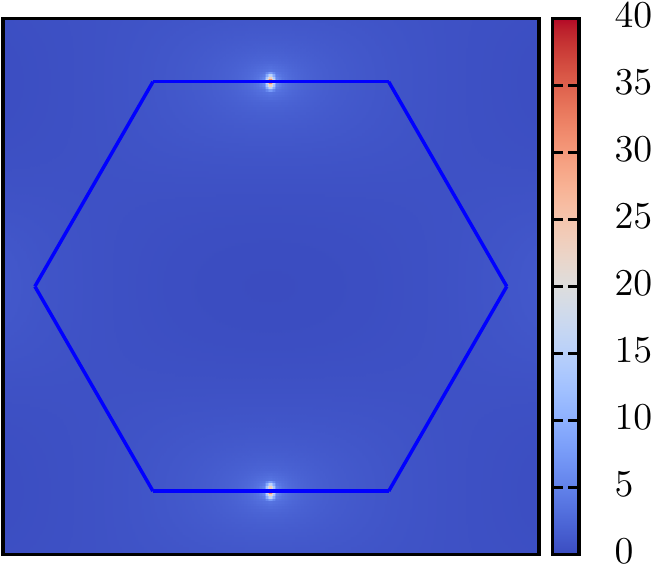}
  \vspace{0.1cm}
  \caption{The static structure factor $S(\vecq)$ for $\kappa=0.61$ and $L=48$: Left: $J_2/J_1 =0$, middle: $J_2/J_1 = 0.18$, right: $J_2/J_1=0.3$. The hexagon is the 1st Brillouin zone of the triangular lattice.}
  \label{stat2}
\end{figure*}
%%%%%%%%%%%%%%%%%%%%%%%%%%%%%%%

Within each of the two phases, a line $\kappa_c(J_2/J_1)$ separates regions of magnetic order ($\kappa>\kappa_c$) and disorder ($\kappa<\kappa_c$). The two methods we have used for determining $\kappa_c$ give lines that track each other closely, but the line from the $m^2$-method is systematically slightly below that from the 
$\xi_a$-method, with the quantitative difference more noticeable in the NSL phase. Our results for $\kappa_c$ for the 0-flux phase should be compared with Ref. \onlinecite{psgWV2006}, where magnetic order
was analyzed using a Bose-Einstein condensation (BEC) approach which involves taking the limit $N\to\infty$ from the outset. While the results appear to agree in the limit $J_2/J_1\to 0$, the difference increases with increasing $J_2/J_1$, with our $\kappa_c$ lying higher. Differences in the predicted onset of magnetic order between finite-$N$ extrapolations and the BEC approach were also noted in Ref. \onlinecite{messio2010}. 

We now turn to the question of the possibility of a spin liquid phase for a certain range of $J_2/J_1$ in the $S=1/2$ model. 
With the conventional identification $\kappa=2S$ in SBMFT this corresponds to $\kappa=1$, which has been considered in previous SBMFT studies of this model.\cite{gazza93,merino} In agreement with these, we find for this case a direct transition at $J_2/J_1\approx 0.16$ between the 120$^{\circ}$ ordered phase and the collinearly ordered phase. On the other hand, referring back to the arguments outlined at the end of Sec. \ref{solving}, it is of interest to also consider smaller values of $\kappa$ as possibly qualitatively relevant for the $S=1/2$ model. Fig. \ref{fig:phasediagram} shows that for $\kappa$ in a small window around $\approx 0.6$, a spin liquid region of the 0-flux type exists for a small range of $J_2/J_1$ values between the two ordered phases. In this $\kappa$ window there are thus two phase transitions as $J_2/J_1$ is increased: a continuous phase transition between the 120$^{\circ}$ ordered phase and the spin liquid, and a first-order transition between the spin liquid and the collinearly ordered phase. The qualitative picture of an intervening spin liquid (without nematic order) between the ordered phases, as well as the nature of the two phase transitions, is in agreement with the findings of Refs. \onlinecite{kaneko,iqbal}. In the SBMFT this scenario arises because the line for $\kappa_c(J_2/J_1)$ 
for the 0-flux state hits the 0-flux/NSL transition line (full black line in Fig. \ref{fig:phasediagram}) at a higher value of $\kappa$ than does the corresponding $\kappa_c$ line for the NSL state. 

To further illustrate the nature of the three phases in this scenario, Fig. \ref{stat2} shows $S(\mathbf{q})$ for $\kappa=0.61$ and three values of $J_2/J_1$, corresponding to representative points within the 120$^{\circ}$ ordered phase, the 0-flux spin liquid state, and the collinearly ordered state. In the 120$^{\circ}$ ordered phase $S(\mathbf{q})$ has very sharp peaks at the Brillouin zone corners. In the spin liquid state the peak locations are the same, but the peaks are considerably lower and broader. In the collinearly ordered state there are again sharp peaks, now located at the midpoint of the horizontal Brillouin zone boundary edge. We note that our plot of $S(\mathbf{q})$ in the spin-liquid phase has the same peak structure as corresponding plots in Refs. \onlinecite{kaneko,iqbal}.

Below this $\kappa$ window the sequence of phases changes. For $\kappa \lesssim 0.57$, a region of NSL opens up between the 0-flux and collinearly ordered state. As $\kappa$ is reduced further, the extent of the two ordered phases diminishes rapidly, with the 120$^{\circ}$ ordered phase disappearing for $\kappa\approx 0.42$. For smaller $\kappa$ the transition to the collinearly ordered state continues to be pushed to higher $J_2/J_1$.\cite{other-phases}       

 % Konklusjon
%%%%%%%%%%%%%%%%%%%%%%%%%%%%%%%%%%%%%%%%%%%%%%%%%%%%%%%%%%%%
\section{Discussion and conclusions}
\label{conclusions}

In this work we have studied the ground state phase diagram of the antiferromagnetic $J_1$-$J_2$ model on the triangular lattice using SBMFT, treating $\kappa=2S$ as a continuous parameter. Motivated by previous numerical and analytical works relevant for the $S=1/2$ case, we have focused our attention on two spin liquid Ans{\"{a}}tze, the 0-flux state\cite{psgWV2006} and a nematic spin liquid (NSL), which upon spinon condensation give rise to, respectively, the two magnetically ordered states known to exist in the $S=1/2$ model, namely the 120$^{\circ}$- ordered state at small $J_2/J_1$ and a collinear stripe ordered state at larger $J_2/J_1$. The need for a nematic spin liquid Ansatz is due to the fact that no symmetric spin liquid Ansatz giving rise to the collinear stripe order has been identified in PSG analyses.

The choice $\kappa=1$ is the standard one for describing the $S=1/2$ model, and for this case we find, in agreement with previous SBMFT studies, a direct, first-order transition between the  two ordered states. We have also explored the phase diagram for $\kappa < 1$, motivated by arguments that in the Schwinger boson mean field theory a smaller value of $\kappa$ may be more appropriate for  qualitatively describing the physics of the exact model. As $\kappa$ is reduced from 1, the first qualitative change in the  the sequence of states (as a function of $J_2/J_1$ for fixed $\kappa$) occurs for $\kappa$ around 0.6, where in a small $\kappa$ window a spin liquid region opens up between the two ordered states. This spin liquid is the 0-flux state. This has several consequences: (i) the static structure factor $S(\mathbf{q})$ of the spin liquid has the same peak locations as in the 120$^{\circ}$-ordered state, (ii) the spin liquid region does not have nematic order, (iii) the transition to the 120$^{\circ}$-ordered state is continuous, (iv) the transition to the collinear stripe state is first-order. We note that these consequences agree with the VMC results of 
Refs. \onlinecite{kaneko} and \onlinecite{iqbal}. On the other hand, these works found gapless spin liquids, while our spin liquid is of the gapped $Z_2$ type.\cite{psgWV2006}  

While a $\kappa$-value as small as $0.73$ for $S=1/2$ can be argued from the requirement that $\langle \mathbf{S}_i^2\rangle$ take its correct value $S(S+1)$, $\kappa$-values as low as $0.6$ are a priori harder to justify. Also, the particular sequence of states only exists in a small kappa window, thus requiring a significant amount of ''fine-tuning.''  To justify our consideration of $\kappa$ values around 0.6, we first note that Ref. \onlinecite{manuel99} calculated 1-loop corrections to SBMFT for $\kappa=1$ and found a magnetically disordered state appearing between the two ordered states. It seems reasonable to guess that this disordered state is the 0-flux spin liquid  found here. If so, it would seem to suggest that the behavior seen in SBMFT for $\kappa  \sim 0.6$ is ''shifted'' to $\kappa\sim 1$ in more accurate calculations that go beyond mean field theory. In fact, a similar conclusion was suggested in an SBMFT study of a different model,\cite{messio2010} namely a nn Heisenberg antiferromagnet perturbed by Dzyaloshinskii-Moriya interactions on the kagome lattice, for which it was found that the SBMFT phase diagram for $\kappa \sim 0.4$ qualitatively resembled exact diagonalization results for the $S=1/2$ model.\cite{cepas2008} We speculate that this might be a quite generic feature of SBMFT: The mean field theory underestimates quantum fluctuations, something which to some extent can be qualitatively compensated for by considering smaller $\kappa$ values, thus giving results that are closer to those of more accurate methods.

We conclude by mentioning some issues that we hope can be resolved in future work. 

As our study based on finite-$N$ calculations and the BEC approach used in Ref. \onlinecite{psgWV2006} give somewhat different predictions for the boundary $\kappa_c(J_2/J_1)$ between magnetic order and disorder in the 0-flux part of the phase diagram (a difference which increases with increasing $J_2$), there is some uncertainty concerning the correct location of such boundaries. We note that a similar comparison for the NSL part of the phase diagram is unavailable as Ref. \onlinecite{psgWV2006} did not consider this state. 

Finally, the possible connection between the NSL state studied here in the bosonic formulation and the nematic spin liquids discussed by Lu\cite{lu1} in the fermionic formulation is not clear to us; it would be interesting to understand this better.

\section*{Acknowledgements}
We acknowledge financial support from NTNU and the Research Council of Norway through its Centres of Excellence funding for ''QuSpin''.

%%%%%%%%%%%%%%%%%%%%%%%%%%%%%%%%%%%%%%%%%%%%%%%%%%%%%%%%%%%%
% Appendix
%%%%%%%%%%%%%%%%%%%%%%%%%%%%%%%%%%%%%%%%%%%%%%%%%%%%%%%%%%%%
\appendix
%%%%%%%%%%%%%%%%%%%%%%%%%%%%%%%%%%%%%%%%%%%%%%%%%%%%%%%%%%%%
% Statisk strukturfaktor
%%%%%%%%%%%%%%%%%%%%%%%%%%%%%%%%%%%%%%%%%%%%%%%%%%%%%%%%%%%%
\section{Static structure factor}
\label{Sqder}
%%%%%%%%%%%%%%%%%%%%%%%%%%%%%%%%%%%%%%%%%%%%%%%%%%%%%%%%%%%%
We will here briefly sketch the derivation of the static structure factor, which is the Fourier transform of the spin-spin correlation function. Using the spin rotation symmetry of the Heisenberg model, and the fact that we work with finite systems so this symmetry is not broken in the ground state, it follows that we can express the spin-spin correlation function as
\begin{eqnarray}
\lefteqn{\braket{\hat{\mathbf{S}}_0\cdot\Si} = 3\braket{\hat{S}_0^z \hat{S}_i^z}} \notag \\
  &=& \frac{3}{4}\left\langle\left(\bzud\bzu-\bzdd\bzd\right)\left(\biud\biu-\bidd\bid\right)\right\rangle.
\end{eqnarray}
As previously, we can introduce Fourier transformed operators to write
\begin{eqnarray} \label{SzSi}
  \lefteqn{\hat{S}_0^z \hat{S}_i^z = \frac{1}{4N^2}\sum_{\veck,\veckp,\vecq,\vecqp}e^{\mathrm{i}(\vecqp-\vecq)\cdot\vecri} \left[\hat{b}_{\veck\uparrow}^\dagger \hat{b}_{\veckp\uparrow}\hat{b}_{\vecq\uparrow}^\dagger \hat{b}_{\vecqp\uparrow}\right.}
  \notag \\ & & \hspace{-1cm}\left.+\hat{b}_{\veck\downarrow}^\dagger \hat{b}_{\veckp\downarrow}\hat{b}_{\vecq\downarrow}^\dagger \hat{b}_{\vecqp\downarrow} 
-\hat{b}_{\veck\uparrow}^\dagger \hat{b}_{\veckp\uparrow}\hat{b}_{\vecq\downarrow}^\dagger \hat{b}_{\vecqp\downarrow}-\hat{b}_{\veck\downarrow}^\dagger \hat{b}_{\veckp\downarrow}\hat{b}_{\vecq\uparrow}^\dagger \hat{b}_{\vecqp\uparrow}\right].
\end{eqnarray}
The expectation values are evaluated by transforming to the basis that diagonalizes the Hamiltonian and using $\hat{\beta}_{\veck\sigma}\ket{\Psi_\text{GS}} = 0$. For example, the first term in \eqref{SzSi} becomes
\begin{eqnarray} 
\lefteqn{\hspace{-1.2cm}\braket{\hat{b}_{\veck\uparrow}^\dagger \hat{b}_{\veck'\uparrow}\hat{b}_{\vecq\uparrow}^\dagger\hat{b}_{\vecq'\uparrow}} 
=\delta_{\veck\vecq'}\delta_{\veck'\vecq} \sinh{\theta_\veck}\cosh{\theta_\veck'}\cosh{\theta_\vecq}\sinh{\theta_{\vecq'}}}\notag\\
&+&\delta_{\veck\veck'}\delta_{\vecq\vecq'}\sinh{\theta_\veck}\sinh{\theta_{\veck'}}\sinh{\theta_\vecq}\sinh{\theta_{\vecq'}}.
\end{eqnarray}
Combining all four terms, after some algebra we get
\begin{eqnarray} 
\lefteqn{\braket{\Sz \cdot \Si} = \frac{3}{2N^2}\sum_{\veck\vecq}\mathrm{e}^{\mathrm{i}(\veck-\vecq)\cdot\vecri}} \notag \\
 &
 \hspace{-0.5cm}\left[(\cosh{2\theta_\veck}-1)(\cosh{2\theta_\vecq+1})-\sinh{2\theta_\veck}\sinh{2\theta_\vecq}\right].
\end{eqnarray}
Invoking the condition \eqref{diagonal-condition} we finally obtain
\begin{equation}
  \braket{\Sz\cdot\Si} = \frac{3}{8N^2}\sum_{\veck, \vecq}\mathrm{e}^{\mathrm{i}(\veck-\vecq)\cdot\vecri}F(\veck, \vecq),
  \label{spincorr-final}
\end{equation}
where
\begin{equation}
F(\veck,\vecq) = \frac{(\gkb+\lambda)(\gqb+\lambda)-\gka\gqa}{\wk\wq}-1.
\end{equation}
This gives the static structure factor 
\begin{equation}
S(\vecq) = \sum_i\braket{\hat{\mathbf{S}}_0\cdot\Si}\mathrm{e}^{-\mathrm{i}\vecq\cdot\vecri} =\frac{3}{8N}\sum_\veck F(\veck,\veck-\vecq).
\end{equation}
We also note that by setting $i=0$ in \eqref{spincorr-final} and using \eqref{mean-field-equations}c and the antisymmetry of $\gamma_k^A$, the result  $\braket{\Si\cdot\Si}=\frac{3}{8}\kappa(\kappa+2)$ follows.

%\newpage

% Liten-kappa analyse
%%%%%%%%%%%%%%%%%%%%%%%%%%%%%%%%%%%%%%%%%%%%%%%%%%%%%%%%%%%%
\section{Small-$\kappa$ analysis}
\label{small-kappa}
%%%%%%%%%%%%%%%%%%%%%%%%%%%%%%%%%%%%%%%%%%%%%%%%%%%%%%%%%%%%

\begin{table*}[htb]\centering
  \begin{tabular}{cccccc}
    \hline
    \multicolumn{2}{r}{0-flux} \\
    \hline\hline
    $i$ & $\lambda_i$ & $a_{1i}$ & $a_{2i}$ & $b_{1i}$ & $b_{2i}$ \\
    \hline
    $0$  & $\frac{1}{4}$ & $\frac{2}{\sqrt{3}}$ & $0$ & $\frac{2}{3}$ & $\frac{2}{3}$   \\
    $1$  & $\frac{1}{24}\cdot\left(11-4j\right)$  & $\frac{\sqrt{3}}{36}\cdot\left(-29+16j\right)$ & $*$ & $-\left(\frac{17}{18}-\frac{8}{3}j\right)$ & $-\left(\frac{11}{2}+\frac{8}{9}j\right)$  \\
    $2$  & $\left(\frac{21935}{13824}+\frac{227}{96}j+\frac{19}{108}j^2\right)$  & $\sqrt{3}\cdot\frac{4993+15840j+2176j^2}{1728}$ & $*$ & $*$ & $*$\\
    \hline
    \multicolumn{2}{r}{NSL} \\
    \hline\hline
    $i$ & $\lambda_i$ & $a_{1i}$ & $a_{2i}$ & $b_{1i}$ & $b_{2i}$ \\
    \hline
    $0$  & $\frac{1}{4}$ & $\sqrt{2}$ & $0$ & $1$ & $1$  \\
    $1$  & $\frac{1}{8}\left(\frac{7}{2}-j\right)$  & $\frac{1}{\sqrt{2}}\cdot\left(-\frac{19}{8}+j\right)$ & $\frac{2\sqrt{2}}{1-j}$ &  $\frac{-13+49j-4j^2}{8(1-j)}$ & $-\left(\frac{5}{8}+\frac{1}{2}j\right)$  \\
    $2$  & $\frac{27-33j+78j^2-24j^3}{128(1-j)}$  & $\sqrt{2}\cdot\frac{-639+1422j+673j^2-752j^3+320j^4}{256(1-j)^2}$ & $*$ & $*$ & $*$ \\
    \hline
  \end{tabular}
\caption{Coefficients needed for the small-$\kappa$ expansion calculation of the ground state energy for the 0-flux and NSL state, up to and including $O(\kappa^3)$ terms. An asterisk (*) indicates either that the coefficient does not appear at this order, or appears in combination with a factor that vanishes, thus giving no contribution. See text for further details.}
\label{small-kappa-table}
\end{table*}

For small values of the parameter $\kappa$, the mean-field equations can be solved by series expansion.\cite{smallkappa} We incorporate the various symmetries of the Ans\"{a}tze directly into the analysis, writing
\begin{subequations}
  \begin{align}
    &\Adi = z_{\delta}^{\Ai} \mathrm{sgn}(\Adn)\Ai, \\
    &\Bdi = z_{\delta}^{\Bi}\Bi,
  \end{align}
\end{subequations}
where $z_\delta^{\Ai/\Bi}=0$ or 1, depending on whether the mean-field parameter vanishes or not. The $\mathrm{sgn}(\Adi)$ takes care of the sign structure of the Ans\"{a}tze, as discussed in section \ref{states-and-ansatze} and summarized in table \ref{ansatztable}. The mean-field equations \eqref{mean-field-equations} can now be rescaled as 
\begin{subequations}
  \begin{align} 
    &\label{MF1}z_{\Ai}\lambda\overset{\sim}{\Ai} = \frac{1}{N}\sum_\veck\frac{{\overset{\sim}{\gamma}_\veck^A}}{\overset{\sim}{\omega}_\veck}\Gamma_\veck^{\Ai}, \\
    &\label{MF2}z_{\Bi}\lambda\overset{\sim}{\Bi} = \frac{1}{N}\sum_\veck\frac{{\overset{\sim}{\gamma}_\veck^B}+1}{{\overset{\sim}{\omega}_\veck}}\Gamma_\veck^{\Bi}, \\
    &\label{MF3}1+\kappa = \frac{1}{N}\sum_\veck\frac{{\overset{\sim}{\gamma}_\veck^A}+1}{{\overset{\sim}{\omega}_\veck}},
  \end{align}
\end{subequations}
where $z_{\Ai/\Bi}=\sum_\delta(z_\delta^{\Ai/\Bi})^2$. In the equations above, $\tilde{O}\equiv O/\lambda$ for $O=\Ai, \Bi, \omega_\veck,\gamma^A_\veck,\gamma_\veck^B$. We have also introduced the Ansatz-dependent factors
\begin{subequations}
  \begin{align}
    &\Gamma_\veck^{\Ai} = \frac{1}{2}\sum_{\delta_i}z_{\delta_i}^{\Ai}\mathrm{sgn}(\Adi)\sin(\veck\cdot\vecd_i),\\
    &\Gamma_\veck^{\Bi} = \frac{1}{2}\sum_{\delta_i}z_{\delta_i}^{\Bi}\cos(\veck\cdot\vecd_i).
  \end{align}
\end{subequations}
The dispersion relation motivates us to expand the mean-field parameters and $\lambda$ in power series as follows:
\begin{subequations} 
  \begin{align}
    &\overset{\sim}{\Ai} = \sqrt{\kappa}\sum_n a_{in}\kappa^n,\\
    &\overset{\sim}{\Bi} = \kappa\sum_n b_{in}\kappa^n,\\
    &\lambda = \sum_n \lambda_n\kappa^n.
  \end{align}
\end{subequations}
We wish to use the small-$\kappa$ expansion to write the ground state energy as
\begin{equation}
\frac{E_0}{J_1 N} = \sum_n e_n \kappa^n.
\end{equation}
This should be compared with 
\begin{align}     \frac{E_0}{J_1 N} &=\frac{1}{2}\sum_{i=1}^2\sum_{\delta_i}\frac{J_i}{J_1}\left({\Bdi}^2-\Adi^2\right) \\
				& = \frac{1}{2}\lambda^2\sum_{i}\frac{J_i}{J_1}\left(z_{\Bi}\Bit-z_{\Ai}\Ait\right).
\end{align}
This gives the following expansion coefficients for the energy:
\begin{align}
& e_0 = 0, \\
& e_1 = -\frac{1}{2}\lambda_0^2(z_{\An}a_{10}^2+jz_{\Ann}a_{20}^2), \\
& e_2 = \frac{1}{2}\lambda_0^2(z_{\Bn}b_{10}^2+jz_{\Bnn}b_{20}^2) - \lambda_0\lambda_1(z_{\An}a_{10}^2+jz_{\Ann}a_{20}^2) \notag \\
	& ~~~~~ -\lambda_0^2(z_{\An}a_{10}a_{11}+jz_{\Ann}a_{20}a_{21}),\\
& e_3 = \frac{1}{2}\lambda_0^2\left[(2z_{\Bn}b_{10}b_{11}-z_{\An}(2a_{10}a_{12}+a_{11}^2))\right.\notag \\
	& ~~~~~~~~~~~~~~\left.+j(2z_{\Bnn}b_{20}b_{21}-z_{\Ann}(2a_{20}a_{22}+a_{21}^2))\right] \notag \\ 
    & ~~~~~+ \lambda_0\lambda_1\left[(z_{\Bn}b_{10}^2-2z_{\An}a_{10}a_{11})\right. \notag \\
    & ~~~~~~~~~~~~~~~~\left.+j(z_{\Bnn}b_{20}^2-2z_{\Ann}a_{20}a_{21})\right] \notag \\
    & ~~~~~- \frac{1}{2}(2\lambda_0\lambda_2+\lambda_1^2)(z_{\An}a_{10}^2+jz_{\Ann}a_{20}^2),
\end{align}
where $j=J_2/{J_1}$~has been introduced. We now expand the mean-field equations \eqref{MF1}-\eqref{MF3} into power series and determine the coefficients $a_{in},b_{in}$ and $\lambda_n$ recursively. To lowest order, only $\lambda_0, a_{10}$ and $a_{20}$ are needed. To obtain the terms up to and including $\mathcal{O}(\kappa^2)$, $\lambda_1,b_{10},b_{20}$ and $a_{11}$ are required. Further including the $\mathcal{O}(\kappa^3)$-terms we additionally need $\lambda_2,b_{11},b_{21},a_{21}$ and $a_{12}$. The values of the coefficients are given in Table \ref{small-kappa-table}. 
For the zero-flux phase, the energy per site is found to be
\begin{align}
\frac{E_{0}^{\text{0-flux}}}{NJ_1} &= -\frac{3}{12}\kappa-\frac{3}{36}\left(\frac{11}{4}-j\right)\kappa^2 \notag \\ &~+\left(-\frac{3}{32}+\frac{19}{72}j-\frac{1}{6}j^2\right)\kappa^3+\mathcal{O}(\kappa^4),
\end{align}
while we for the NSL phase obtain
\begin{align}
\frac{E_{0}^{\text{NSL}}}{NJ_1} &= -\frac{3}{12}\kappa-\frac{3}{36}\left(\frac{21}{8}-\frac{3}{4}j\right)\kappa^2 \notag \\ &~-\frac{9-25j+26j^2-8j^3}{128(1-j)}\kappa^3+\mathcal{O}(\kappa^4).
\end{align}
Thus, the energy difference is
\begin{eqnarray}
\lefteqn{\frac{E_{0}^{\text{NSL}}-E_{0}^{\text{0-flux}}}{NJ_1} = \left(\frac{3}{8}-\frac{3}{4}j\right)\left(\frac{\kappa}{6}\right)^2} \notag \\ &+&\left(\frac{-27+187j-262j^2+120j^3}{32}\right)\left(\frac{\kappa}{6}\right)^3.
\end{eqnarray}  
The phase transition line $E_{0}^{\text{NSL}}=E_{0}^{\text{0-flux}}$ is plotted from this expression as a dashed line in Fig. \ref{fig:phasediagram}. One sees that it agrees with our numerical results for small enough $\kappa$, approaching 
$j=1/2$ in the limit $\kappa \to 0$.

%%%%%%%%%%%%%%%%%%%%%%%%%%%%%%%%%%%%%%%%%%%%%%%%%%%%%%%%%%%%
% Referanser
%%%%%%%%%%%%%%%%%%%%%%%%%%%%%%%%%%%%%%%%%%%%%%%%%%%%%%%%%%%%
%\bibliographystyle{unsrt}
%\bibliography{mybib.bib}

\begin{thebibliography}{0}%
\makeatletter
\providecommand \@ifxundefined [1]{%
 \@ifx{#1\undefined}
}%
\providecommand \@ifnum [1]{%
 \ifnum #1\expandafter \@firstoftwo
 \else \expandafter \@secondoftwo
 \fi
}%
\providecommand \@ifx [1]{%
 \ifx #1\expandafter \@firstoftwo
 \else \expandafter \@secondoftwo
 \fi
}%
\providecommand \natexlab [1]{#1}%
\providecommand \enquote  [1]{``#1''}%
\providecommand \bibnamefont  [1]{#1}%
\providecommand \bibfnamefont [1]{#1}%
\providecommand \citenamefont [1]{#1}%
\providecommand \href@noop [0]{\@secondoftwo}%
\providecommand \href [0]{\begingroup \@sanitize@url \@href}%
\providecommand \@href[1]{\@@startlink{#1}\@@href}%
\providecommand \@@href[1]{\endgroup#1\@@endlink}%
\providecommand \@sanitize@url [0]{\catcode `\\12\catcode `\$12\catcode
  `\&12\catcode `\#12\catcode `\^12\catcode `\_12\catcode `\%12\relax}%
\providecommand \@@startlink[1]{}%
\providecommand \@@endlink[0]{}%
\providecommand \url  [0]{\begingroup\@sanitize@url \@url }%
\providecommand \@url [1]{\endgroup\@href {#1}{\urlprefix }}%
\providecommand \urlprefix  [0]{URL }%
\providecommand \Eprint [0]{\href }%
\providecommand \doibase [0]{http://dx.doi.org/}%
\providecommand \selectlanguage [0]{\@gobble}%
\providecommand \bibinfo  [0]{\@secondoftwo}%
\providecommand \bibfield  [0]{\@secondoftwo}%
\providecommand \translation [1]{[#1]}%
\providecommand \BibitemOpen [0]{}%
\providecommand \bibitemStop [0]{}%
\providecommand \bibitemNoStop [0]{.\EOS\space}%
\providecommand \EOS [0]{\spacefactor3000\relax}%
\providecommand \BibitemShut  [1]{\csname bibitem#1\endcsname}%
\let\auto@bib@innerbib\@empty
%</preamble>
\end{thebibliography}%


\begin{references}

\bibitem{bal10} L. Balents,  Nature \textbf{464}, 199 (2010).

\bibitem{mis-lhu} G. Misguich and C. Lhuillier, in ''Frustrated spin systems'', edited by H. T. Diep (World Scientific, 2nd ed., 2013), also available as arXiv:cond-mat/0310405v2.

\bibitem{sav-bal} L. Savary and L. Balents, arXiv:1601.03742.

\bibitem{norman} For a recent review, see M. Norman, Rev. Mod. Phys. \textbf{88}, 041002 (2016).

\bibitem{LSM-16} A. M. L\"{a}uchli, J. Sudan, and R. Moessner, arXiv:1611.06990v1.

\bibitem{anderson73} P. W. Anderson, Mater. Res. Bull. \textbf{8}, 153 (1973).

\bibitem{trlatt-order} D. A. Huse and V. Elser, Phys. Rev. Lett. \textbf{60}, 2531 (1988); T. Jolicoeur and J. C. Le Guillou, Phys. Rev. B \textbf{40}, 2727(R) (1989); B. Bernu, C. Lhuillier, and L. Pierre, Phys. Rev. Lett. \textbf{69}, 2590 (1992); L. Capriotti, A. E. Trumper, and S. Sorella, Phys. Rev. Lett. \textbf{82}, 3899 (1999).

\bibitem{mishmash} R. V. Mishmash, J. R. Garrison, S. Bieri, and C. Xu, Phys. Rev. Lett. \textbf{111}, 157203 (2013). 

\bibitem{kaneko} R. Kaneko, S. Morita, and M. Imada, J. Phys. Soc. Jpn. \textbf{83}, 093707 (2014).

\bibitem{li} P. H. Y. Li, R. F. Bishop, and C. E. Campbell, Phys. Rev. B \textbf{91}, 014426 (2015); R. F. Bishop and P. H. Y. Li, Europhys. Lett. \textbf{112}, 67002 (2015).

\bibitem{white} Z. Zhu and S. R. White, Phys. Rev. B \textbf{92}, 041105(R) (2015).

\bibitem{sheng1} W.-J. Hu, S.-S. Gong, W. Zhu, and D. N. Sheng, Phys. Rev. B 92, 140403(R) (2015).

% Classical and semiclassical studies

\bibitem{kors93} S. E. Korshunov, Phys. Rev. B \textbf{47}, 6165(R) (1993).

\bibitem{jol90} T. Jolicoeur, E. Dagotto, E. Gagliano, and S. Bacci, Phys. Rev. B \textbf{42}, 4800 (1990).

\bibitem{messio11} L. Messio, C. Lhuillier, and G. Misguich, Phys. Rev. B \textbf{83}, 184401 (2011).

\bibitem{chub92} A. V. Chubukov and T. Jolicoeur, Phys. Rev. B \textbf{46}, 11137 (1992).

\bibitem{iv93} N. B. Ivanov, Phys. Rev. B \textbf{47}, 9105 (1993).

\bibitem{ritchey90} I. Ritchey, P. Chandra, and P. Coleman, Phys. Rev. Lett. \textbf{64}, 2583 (1990).

\bibitem{deutscher93} R. Deutscher and H. U. Everts, Z. Phys. B \textbf{93}, 77 (1993).

% Recent numerical studies

\bibitem{manuel99} L. O. Manuel and H. A. Ceccatto, Phys. Rev. B \textbf{60}, 9489 (1999).

\bibitem{iqbal} Y. Iqbal, W.-J. Hu, R. Thomale, D. Poilblanc, and F. Becca, Phys. Rev. B \textbf{93}, 144411 (2016).

\bibitem{sheng2} W.-J. Hu, S.-S. Gong, and D. N. Sheng, Phys. Rev. B \textbf{94}, 075131 (2016).

\bibitem{ian} S. N. Saadatmand and I. P. McCulloch, Phys. Rev. B \textbf{94}, 121111(R) (2016).

\bibitem{lauchli} A. Wietek and A. M. L\"{a}uchli, Phys. Rev. B \textbf{95}, 035141 (2017).

\bibitem{ian2} S.N. Saadatmand and I. McCulloch, Phys. Rev. B \textbf{96}, 075117 (2017).

\bibitem{sheng3} S.-S. Gong, W. Zhu, J.-X. Zhu, D. N. Sheng, and K. Yang, Phys. Rev. B \textbf{96}, 075116 (2017).

\bibitem{caveats} Due to features in the entanglement spectrum, Ref. \onlinecite{ian} was unable to rule out that the system becomes gapless in the limit of infinite cylinder width. Ref. \onlinecite{sheng3} found that the spin triplet gap in the even topological sector vanishes (in contrast to the odd sector where it is large).

% Schwinger boson and Abrikosov fermion studies

\bibitem{arovas-auerbach} D. P. Arovas and A. Auerbach, Phys. Rev. B \textbf{38}, 316 (1988); A. Auerbach and D. P. Arovas, Phys. Rev. Lett. \textbf{61}, 617 (1988). 

\bibitem{gazza93} C. J. Gazza and H. A. Ceccatto, J. Phys.: Condens. Matter \textbf{5}, L135 (1993). 

\bibitem{merino} J. Merino, M. Holt, and B. J. Powell, Phys. Rev. B \textbf{89}, 245112 (2014).

\bibitem{messio2010} L. Messio, O. C\'{e}pas, and C. Lhuillier, Phys. Rev. B \textbf{81}, 064428 (2010). 

\bibitem{psgmessio2013} L. Messio, C. Lhuillier, and G. Misguich, Phys. Rev. B \textbf{87}, 125127 (2013).

\bibitem{psgwen2002prb65} X.-G. Wen, Phys. Rev. B \textbf{65}, 165113 (2002).
 
\bibitem{wenbook} X.-G. Wen, \textit{Quantum field theory of many-body systems} (Oxford University Press, 2004).

\bibitem{psgWV2006} F. Wang and A. Vishwanath, Phys. Rev. B \textbf{74}, 174423 (2006).

\bibitem{qi1} W. Zheng, J.-W. Mei, and Y. Qi, arXiv:1505.05351.

\bibitem{lu1} Y.-M. Lu, Phys. Rev. B \textbf{93}, 165113 (2016).

\bibitem{qi2} Y. Qi and M. Cheng, arXiv:1606.04544.

\bibitem{lu2} Y.-M. Lu, arXiv:1606.05652.

\bibitem{sachdev92} S. Sachdev,  Phys. Rev. B \textbf{45}, 12377 (1992).

\bibitem{terminology} This is the terminology used in Ref. \onlinecite{psgWV2006}, which also
characterized the Ans{\"{a}}tze as triples $(p_1,p_2,p_3)$ with $p_i=0,1$. The 0-flux state (0,0,1) is called B5 in Ref. \onlinecite{qi1} and \#20 in Ref. \onlinecite{lu1}. The $\pi$-flux state (1,1,0) is called A1 in Ref. \onlinecite{qi1} and \#1 in Ref. \onlinecite{lu1}.

\bibitem{pifluxcriticism} We note that a later SBMFT PSG analysis by Messio et al.\cite{psgmessio2013} (which also included chiral states) did not find the $\pi$-flux state among the nonchiral solutions and argued against its existence.

\bibitem{auerbach-book} A. Auerbach, \textit{Interacting electrons and quantum magnetism} (Springer-Verlag, Berlin, 1994).

%%%%%%

\bibitem{compare-mezio} We note that the expressions in (\ref{magnetization-parameter}), specialized to the 120$^{\circ}$ ordered state, differ slightly from corresponding expressions given in Ref. \onlinecite{mezio}.

\bibitem{mezio} A. Mezio, C. N. Sposetti, L. O. Manuel, and A. E. Trumper, Europhys. Lett. \textbf{94}, 47001 (2011); A. Mezio, L. O. Manuel, R. R. P. Singh, and A. E. Trumper, New J. Phys. \textbf{14}, 123033 (2012).

\bibitem{fits} In the fits we used the exact (middle) expression in (\ref{magnetization-parameter}). We also verified numerically that in the ordered phases the ratio between the approximate (rightmost) and exact expression increases towards 1 as $N$ is increased.

\bibitem{sandvik-scaling-1997} A. W. Sandvik, Phys. Rev. B \textbf{56}, 11678 (1997).

\bibitem{sandvik-correlation-2010} A. W. Sandvik, AIP Conf. Proc. \textbf{1297}, 135 (2010); also available as arXiv:1101.3281.

\bibitem{smallkappa} O. Tchernyshyov, R. Moessner, and S. L. Sondhi, Europhys. Lett. \textbf{73}, 278 (2006).

\bibitem{other-phases} For sufficiently large $J_2/J_1$ we expect that other phases than NSL, not considered in our analysis, will become competitive. This will also depend on $\kappa$. For example, for $\kappa=1$, Ref. \onlinecite{gazza93} found a transition to a phase with incommensurate spiral order for $J_2/J_1\approx 0.95$. 

\bibitem{cepas2008} O. C\'{e}pas, C. M. Fong, P. W. Leung, and C. Lhuillier, Phys. Rev. B \textbf{78}, 140405(R) (2008).

\end{references}

\end{document}